\documentclass[preprint]{vgtc}               % preprint
\ifpdf%                                % if we use pdflatex
  \pdfoutput=1\relax                   % create PDFs from pdfLaTeX
  \usepackage{graphicx}                % allow us to embed graphics files
  \DeclareGraphicsExtensions{.pdf,.png,.jpg,.jpeg} % for pdflatex we expect .pdf, .png, or .jpg files
\else%                                 % else we use pure latex
  \usepackage{graphicx}                % allow us to embed graphics files
  \DeclareGraphicsExtensions{.eps}     % for pure latex we expect eps files
\fi%

\graphicspath{{results/}} % where to search for the images

\usepackage{microtype}                 % use micro-typography (slightly more compact, better to read)
\PassOptionsToPackage{warn}{textcomp}  % to address font issues with \textrightarrow
\usepackage{textcomp}                  % use better special symbols
\usepackage{mathptmx}                  % use matching math font
\usepackage{times}                     % we use Times as the main font
\usepackage{cite}                      % needed to automatically sort the references
\usepackage{tabu}                      % only used for the table example
\usepackage{booktabs}                  % only used for the table example
\usepackage{amsmath,amsfonts,amssymb}
\usepackage{subfig}
\usepackage[group-separator={,}]{siunitx}

\ifdefined\ShowRevision
\usepackage{color}
\usepackage[normalem]{ulem}

\newcommand{\revi}[1] {{\color{blue}#1}}
\newcommand{\revirm}[1] {\revi{\sout{#1}}}
\else
\newcommand{\revi}[1] {#1}
\newcommand{\revirm}[1] {}
\fi

\onlineid{1052}

\vgtccategory{Research}

\vgtcinsertpkg

\newcommand{\R}{\mathbb{R}}

\DeclareMathOperator{\trace}{tr}
\newcommand{\Identity}{I}

\newcommand{\dimension}{n}

\newcommand{\vf}{v}
\newcommand{\flowmap}{\phi}
\newcommand{\straintensor}{\mathbb{C}}
\DeclareMathOperator{\ftle}{FTLE}

\newcommand \dd[1]  { \,\textrm d{#1}}

\newcommand{\Wiener}{W}
\newcommand{\Disturbance}{B}
\newcommand{\Diffusion}{D}
\newcommand{\scaling}{s}

\newcommand{\avgstrain}{\mathbb{\bar{C}}}

\DeclareMathOperator{\DBS}{DBS}

\newcommand{\Cov}{C}
\newcommand{\CovSet}{\mathcal{C}}

\DeclareMathOperator{\Expectation}{\mathbb{E}}
\newcommand{\mean}{\mu}
\newcommand{\std}{\sigma}

\newcommand{\tv}{\sigma_T}

\newcommand{\ms}[1]{\num{#1}\si{\milli\second}}

\title{Uncertain Transport in Unsteady Flows}

\author{Tobias Rapp\thanks{e-mail: tobias.rapp@kit.edu} %
	\and Carsten Dachsbacher\thanks{e-mail: dachsbacher@kit.edu}}
\affiliation{\scriptsize Institute for Visualization and Data Analysis \\ Karlsruhe Institute of Technology}
\teaser{
	\centering
	\subfloat[Backward diffusion barrier strength (DBS)]{%
		\includegraphics[width=0.45\linewidth]{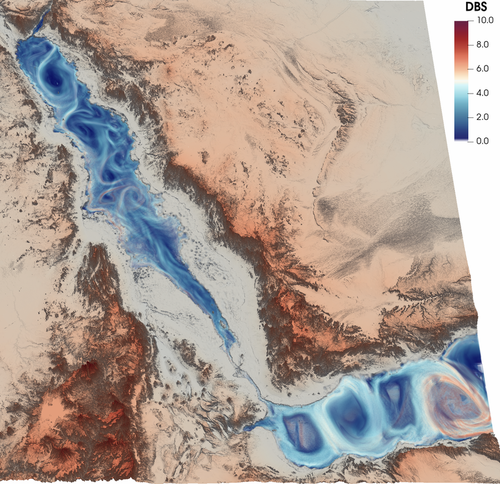}%
	}
	\hfill
	\subfloat[Transport uncertainty]{%
		\includegraphics[width=0.45\linewidth]{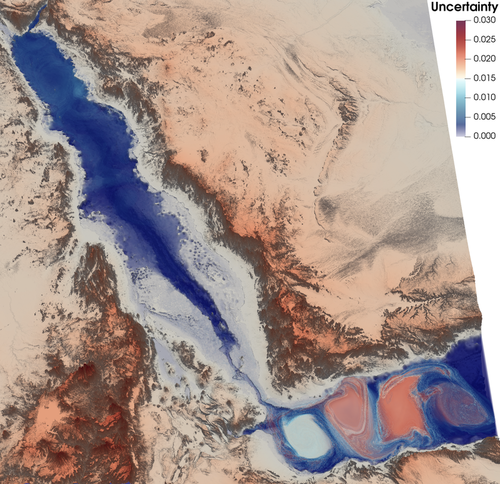}%
	}
	\vspace*{-1mm}
	\caption{We visualize transport under uncertainties in the Red Sea ensemble dataset.
	The backward DBS in (a) indicates material surfaces that are maximally diffusive.
	Since the DBS assumes only small-scale stochastic deviations, we propose a complementary visualization of the absolute scale of uncertainties in the Lagrangian frame (b).
	}
  \label{fig:teaser}
}

\abstract{
	We study uncertainty in the dynamics of time-dependent flows by identifying barriers and enhancers to stochastic transport. This topological segmentation is closely related to the theory of Lagrangian coherent structures and is based on a recently introduced quantity, the diffusion barrier strength (DBS).
	The DBS is defined similar to the finite-time Lyapunov exponent (FTLE), but incorporates diffusion during flow integration.
	Height ridges of the DBS indicate stochastic transport barriers and enhancers, i.e.\ material surfaces that are minimally or maximally diffusive.
	To apply these concepts to real-world data, we represent uncertainty in a flow by a stochastic differential equation that consists of a deterministic and a stochastic component modeled by a Gaussian.
	With this formulation we identify barriers and enhancers to stochastic transport, without performing expensive Monte Carlo simulation and with a computational complexity comparable to FTLE\@.
	In addition, we propose a complementary visualization to convey the absolute scale of uncertainties in the Lagrangian frame of reference.
	This enables us to study uncertainty in real-world datasets, for example due to small deviations, data reduction, or estimated from multiple ensemble runs.
} % end of abstract

\CCScatlist{
  \CCScatTwelve{Human-centered computing}{Visu\-al\-iza\-tion}{Visu\-al\-iza\-tion application domains}{Scientific visualization};
}

\begin{document}

%% The ``\maketitle'' command must be the first command after the
%% ``\begin{document}'' command. It prepares and prints the title block.

%% the only exception to this rule is the \firstsection command
\firstsection{Introduction}

\maketitle

%% \section{Introduction} %for journal use above \firstsection{..} instead
Although most experiments and simulations produce deterministic data, uncertainty exists in all measured or simulated flows.
This uncertainty might be estimated from repeated simulation runs or measurements, it might be introduced by data processing and reduction, or it can be explicitly modeled.
Studying uncertainty is especially relevant in unsteady flows, where small variations in the initial conditions can cause dramatic changes to the flow.
%TODO More Uncertainty vis.!?
In this paper, we investigate uncertainties in the Lagrangian transport, i.e.\ the advection of a material by the flow.

% LCS
For deterministic flows, the Lagrangian coherent structures (LCS) identify a topological skeleton of the flow dynamics in a finite-time interval.
%LCS are material surfaces that remain coherent over time, although different definitions of coherency exist~\cite{Hadjighasem2017}.
%These coherent structures form barriers or enhancers to the flow and thus determine the global transport and mixing behavior.
%
% Probabilistic LCS
Recent work has extended the definitions of coherent structures to uncertain flows. The probabilistic~\cite{Guo2016} or averaged~\cite{Schneider2012} transport is estimated using a Monte Carlo approach, i.e.\ by advecting a large amount of particles.
%Different approaches to estimate the transport behavior, such as probabilistic~\cite{Guo2016} or averaged~\cite{Schneider2012} quantities have been proposed.
While the LCS are theoretically well established, this is, to our knowledge, not the case for its probabilistic extension.

% DBS  
Based on recent work from Haller, Karrasch, and Kogelbauer~\cite{Karrasch2018, Karrasch2020}, we employ the diffusion barrier strength (DBS) to identify transport barriers and enhancers to stochastic flows. These are material surfaces that show minimal or maximal stochastic cross flux. % (wrt small-scale deviations)
By assuming only small stochastic deviations, Monte Carlo integration is avoided and only the deterministic part of the flow has to be advected.
To this end, we first define uncertain unsteady flows as stochastic differential equations that consist of an advective component and an added stochastic component modeled as a Gaussian.
The central limit theorem makes this assumption reasonable.
In this paper, we discuss how to model uncertainty information in this stochastic differential equation, e.g. due to data reduction, to model small-scale deviations, or to model aggregated ensemble members.
To complement the visualization of stochastic transport barriers and enhancers, which is based on the assumption of small-scale deviations, we propose a novel visualization of the scale of uncertainties encountered during advection.
In several experiments, we investigate the relationship between the stochastic transport barriers and enhancers, Lagrangian coherent structures, and its probabilistic extensions. 

%\pagebreak
\noindent To summarize, our contributions are:
\begin{itemize}
	\item We model stochastic flows with small deviations to Gaussian flow fields~(\autoref{sec:Modeling}),
	\item We propose a novel visualization of transport uncertainties~(\autoref{sec:TransportUncertainty}),
	\item We apply the theory of stochastic transport barriers and enhancers to real-world data and compare it to probabilistic extensions of the LCS~(\autoref{sec:Results}).
\end{itemize}

\section{Related Work}
\label{sec:RelatedWork}

Uncertainty visualization has been an active research topic in the field of visualization for more than two decades~\cite{Johnson2003}.
Several surveys~\cite{Pang1997, Bonneau2014} motivate uncertainty visualization, introduce its challenges, and different sources of uncertainty.
In this study, we focus on the visualization of uncertain and unsteady flows. We recapitulate the theory of Lagrangian coherent structures before we discuss uncertain Lagrangian approaches.

%\paragraph*{Uncertain Flow Visualization}

%For time-dependent 2D datasets, texture-based methods have been proposed by Botchen et al.~\cite{Botchen2005} and Osorio and Brodlie~\cite{Osorio2009}. In contrast, Hlawatsch et al.~\cite{Hlawatsch2011} visualize uncertainty in unsteady flows by introducing flow radar glyphs.
%
%Otto et al.~\cite{Otto2010, Otto2011closed, Otto2011} introduce vector field topology for uncertain steady flows. Otto and Theisel~\cite{Otto2012vortex} study the extraction and visualization of vortices in uncertain vector fields using a Monte Carlo approach. Similarly, Petz et al.~\cite{Petz2012} use Monte Carlo estimation in Gaussian random fields to extract probabilistic local features.
%Ferstl et al.~\cite{Ferstl2016} visualize the uncertainty in vector field ensembles by clustering streamlines of ensemble members and by visualizing the confidence regions.
%Bhatia et al.~\cite{Bhatia2012} analyze and quantify the uncertainty introduced by streamline integration, whilst Chen et al.~\cite{Chen2015} discuss and model the error introduced by integrating pathlines in unsteady flows.

\paragraph*{Lagrangian Coherent Structures}

LCS are attracting and repelling material surfaces that separate regions of different flow behavior in a finite-time interval~\cite{Haller2015}. Since these material surfaces show minimal or maximal cross flux, they control the global transport and mixing behavior~\cite{Shadden2005, Mathur2007}.
Due to differing views on coherency, several approaches exist to characterize LCS~\cite{Hadjighasem2017}, each of which may lead to different results.
The finite-time Lyapunov exponent (FTLE) measures the separation or attraction of infinitesimally close tracer particles and is closely related to the LCS~\cite{Haller2000}, which can be defined as height ridges of the FTLE\@. The extraction of LCS as height ridges has been investigated by Sadlo and Peikert~\cite{Sadlo2007} and subsequent studies~\cite{Schindler2012, Rapp2019vmv}.
Although research on the identification of LCS is still ongoing, the FTLE has been established as a powerful visualization of the time-dependent flow dynamics. The efficient computation of the FTLE has been an active research area since it requires a dense integration of tracer particles~\cite{Garth2007, Kuhn2012}. In general, to compute the FTLE in an $\dimension$-dimensional time-dependent flow $\vf(x, t)$, a particle at position $x_0 \in \R^{\dimension}$ at time $t_0$ is advected to time $t_1$ by solving an ordinary differential equation
\begin{equation}
\label{eq:ODE}
\dd x(t) = \vf(x(t), t) \dd t, \qquad x(t_0) = x_0.
\end{equation}
The FTLE is computed from the flow map $\flowmap(x, t_0, t_1)$, which maps a position $x$ at time $t_0$ to a position at time $t_1$. From the spatial gradient $\nabla \flowmap$, the right Cauchy-Green strain tensor is defined as:
\begin{equation}
\straintensor(x, t_0, t_1) := \nabla \flowmap(x, t_0, t_1)^\top \nabla \flowmap(x, t_0, t_1).
\end{equation}
The FTLE is then computed using the largest eigenvalue $\lambda_{\max}$ of the strain tensor:
\begin{equation}
\ftle(x, t_0, t_1) := \frac{1}{|t_1 - t_0|} \log\sqrt{\lambda_{\max}( \straintensor(x, t_0, t_1)) }.
\end{equation}
The FTLE thus describes the average exponential stretching of an infinitesimally close volume at time $t_0$ when the flow is integrated to $t_1$.

\paragraph*{Uncertain Lagrangian Transport}

In uncertain flows, particles are advected stochastically. The flow map thus describes a distribution of positions where particles might be advected.
Schneider et al.~\cite{Schneider2012} estimate this stochastic flow map using a Monte Carlo approach. The authors then estimate the variance in the stochastic flow map, which defines the finite-time variance analysis (FTVA), a FTLE-like metric.
Hummel et al.~\cite{Hummel2013} discuss the comparative visual analysis of Lagrangian transport in CFD ensembles based on the FTVA\@.
Guo et al.~\cite{Guo2016} propose two extensions of the FTLE: by estimating the expectation of the strain tensor and then computing a single FTLE value (FTLE-D), or by estimating a distribution of FTLEs (D-FTLE). Both approaches depend on Monte Carlo estimation of the stochastic flow~\cite{Guo2019}. In this study, we present a new quantity that does not require expensive Monte Carlo estimation and is built upon a more solid theoretical foundation.

% Hollister et al.~\cite{Hollister2016} cluster ensemble members to investigate similarities in the transport. ?
% Jarema et al. "Comparative Visual Analysis of Transport Variabilityin Flow Ensembles" ?

\section{Stochastic Flows}
\label{sec:SDE}

To visualize uncertainty in the transport in unsteady flows, we first introduce a stochastic flow as a deterministic flow with small stochastic deviations. More formally, we model an uncertain flow by a stochastic differential equation (SDE), i.e.\ we extend the ordinary differential equation from \autoref{eq:ODE} with a stochastic component
\begin{equation}
\label{eq:SDE}
\dd x(t) = \underbrace{\vf(x(t), t) \dd t}_{\text{deterministic}} + \underbrace{\sqrt{\scaling} \Disturbance(x(t), t) \dd \Wiener(t)}_{\text{stochastic}}.
\end{equation}
Here, $\Wiener(t)$ is an $\dimension$-dimensional Wiener process with disturbance $\sqrt{\scaling} \Disturbance(x(t), t)$.
The Wiener process $\Wiener$ consists of independent standard Gaussian distributions at every time $t$. The notation $\dd \Wiener(t)$ represents a random variable that is distributed with respect to a standard, multivariate Gaussian.
The disturbance, which controls the scaling and anisotropy, is separated into a scaling parameter $\scaling > 0$ and a scale-independent matrix $\Disturbance \in \R^{\dimension \times \dimension}$. In the following, we will assume only small deviations, i.e.~$\scaling$ is small.

\paragraph*{Numerical Integration}
\label{sec:SDE:Integration}

In general, SDEs can be solved by numerical integration using e.g.\ the Euler-Marayuma or the Runge-Kutta methods for SDEs~\cite{Kloeden}. These Markov chain Monte Carlo strategies involve sampling of the stochastic component. The numerical integration is thus significantly more involved compared to deterministic flows since it requires a large amount of stochastically integrated particles. At the same time, it is non-trivial to decide how many particles should be integrated. For these reasons, we want to avoid the numerical integration of stochastic flows.

%Guo et al.~\cite{Guo2016} propose to stop the computation if the posterior distribution does not statistically significantly change anymore. However, it is possible that this strategy misses significant flow behavior, e.g.~due to strong non-linearities in the flow. This can only be addressed by integrating more particles.

\section{Stochastic Transport Barriers and Enhancers}
\label{sec:DBS}

In this section, we introduce stochastic transport barriers and enhancers. We focus on an intuitive introduction and refer to the work of Haller, Karrasch, and Kogelbauer for the formal derivation~\cite{Karrasch2018, Karrasch2020}. 
Transport barriers are inhibitors of the spread of substances in a flow, whilst transport enhancers maximize such diffusion or mixing processes. Remarkably, these barriers and enhancers do not depend on the actual value of the diffusivities, i.e.\ the scaling parameter $s$.
They are also well-defined for deterministic flows when we consider the case of $s \rightarrow 0$.
In this case, they present an alternative to the Lagrangian coherent structures, but do not depend on any specific definition of coherency.
%
%In practice, the transport barriers and enhancers can only be analytically determined for two-dimensional flows~\cite{Karrasch2018}.
The diffusion barrier strength (DBS) visualizes the barriers and enhancers, which can be defined as ridges of the DBS, similar to the LCS that can be defined as ridges of the FTLE\@.

The DBS is computed from a deterministic flow $\vf$ and a diffusion component that describes the amount and anisotropy of diffusion at each point in space and time. First, we introduce the tensor $T$ from the gradient of the flow map $\nabla \flowmap$ and the diffusion $\Diffusion \in \R^{\dimension \times \dimension}$  as
\begin{equation}
\label{eq:transport_tensor}
T(x, t_0, t) := \left[ \nabla \flowmap(x, t_0, t) \right]^{-1} D(x, t) \left[ \nabla \flowmap(x, t_0, t) \right]^{-T}.
\end{equation}
If the diffusion is isotropic, i.e.\ $D \equiv \Identity$, then
\begin{equation}
T(x, t_0, t) = \straintensor(x, t_0, t)^{-1}.
\end{equation}
However, we have to incorporate the diffusion $\Diffusion(x, t)$ at every time in the interval $[t_0, t_1]$, in contrast to the FTLE that only considers the deformation at the end of the time interval.
Therefore, the time-averaged, diffusivity-weighted right Cauchy-Green strain tensor $\avgstrain$ is computed as
\begin{equation}
\label{eq:AvgStrain}
\avgstrain(x_0, t_0, t_1) := \frac{1}{|t_1 - t_0|} \int_{t_0}^{t_1} \det(\Diffusion(x, t)) T(x, t_0, t)^{-1} \dd t,
\end{equation}
where $x$ is the position during integration at time $t$, i.e.\ $x = \flowmap(x_0, t_0, t)$. Since we need only the inverse of $T$, we compute
\begin{equation}
\label{eq:inv_transport_tensor}
T(x, t_0, t)^{-1} = \left[ \flowmap(x, t_0, t)\right]^\intercal D(x, t)^{-1} \left[\nabla \flowmap(x, t_0, t)\right]
\end{equation}
instead of \autoref{eq:transport_tensor}.
Lastly, Haller et al.~\cite{Karrasch2018} define the DBS as the trace of $\avgstrain$. Since this quantity is exponential, we take the logarithm \revi{for} visualization:
\begin{equation}
\DBS(x_0, t_0, t_1) := \log \left( \trace(\avgstrain(x_0, t_0, t_1)) \right).
\end{equation}
Although the integral in \autoref{eq:AvgStrain} might seem daunting at first, we are already performing this integration when computing the flow map $\flowmap$. Thus, to compute the DBS, we integrate the deterministic flow $\vf$ and at each step evaluate $T^{-1}$ to accumulate the diffusivity-weighted and time-averaged strain tensor $\avgstrain$.

\section{Modeling Diffusion}
\label{sec:Modeling}

To compute the DBS, we require a scale-independent diffusion component $\Diffusion$.
For completely deterministic flows, we set the diffusion to the identity matrix, i.e.\ $\Diffusion = \Identity$. For stochastic flows, with a scale-independent disturbance $\Disturbance$ (cf.\ \autoref{eq:SDE}), the diffusion is defined as
\begin{equation}
\Diffusion = \frac{1}{2}\Disturbance \Disturbance^\intercal.
\end{equation}
For uncertain and unsteady flows modeled by Gaussians, we now discuss how to obtain $\Disturbance$. The scale-independent disturbance represents the anisotropy and the scaling relative to other regions of the flow.
Given a Gaussian with covariance $\Cov(x, t)$, we want to separate it into a global scaling parameter $\scaling$ and a disturbance $\Disturbance(x, t)$.

Since the disturbance should be, on average, centered around the identity matrix $\Identity$, we standardize all covariance matrices. That is, given the set of all covariance matrices $\CovSet$, we subtract the mean of all variances, i.e.\ the diagonal elements of each covariance matrix $\Cov \in \CovSet$. Then, we divide out the maximal standard deviation over all dimensions:
\begin{equation}
\Disturbance(x, t) = \frac{\Cov(x, t) - \Identity \mean_\Cov}{\std^{\max}_{\Cov}},
\end{equation}
where $\mean_\Cov$ is the mean of all variances:
\begin{equation}
\mean_\Cov := 
\begin{bmatrix}
\Expectation[\CovSet_{0, 0}] \\
\vdots \\
\Expectation[\CovSet_{n-1, n-1}]
\end{bmatrix}
\end{equation}
and $\std^{\max}_{\Cov}$ is the maximum of the standard deviation of all variances in $\CovSet$:
\begin{equation}
\std^{\max}_{\Cov} := \max\left(\sqrt{\Expectation[\CovSet_{i, i} - \mean_{\Cov_{i, i}}]}\right), \quad \text{where } i = 0 \dots n-1.
\end{equation}
Although a different scaling than $\std^{\max}_{\Cov}$ could be used since it is canceled out in \autoref{eq:AvgStrain}, our definition increases the numerical stability.

\section{Visualizing Transport Uncertainty}
\label{sec:TransportUncertainty}

By design, the diffusion barrier strength ignores the absolute scale of stochasticity, i.e.\ the amount of uncertainty of the transport. However, this quantity is still relevant, especially if the amount of stochastic deviations varies strongly in the flow.
To this end, we propose a visualization that complements the DBS by directly conveying the scale of stochastic deviations.

% Material transport of uncertainty
	
Although it is possible to directly visualize the time-dependent variance of a Gaussian flow field, we are interested in the uncertainty of the transport, which is inherently defined in a Lagrangian frame.
We propose to measure the uncertainty encountered during the integration of a tracer particle. In other words, this visualizes the transport of uncertainty in the flow.
Moreover, this enables us to integrate only the deterministic part of the stochastic flow and avoid stochastic numerical integration.

First, we discuss how to measure the uncertainty of a Gaussian flow with covariance $\Cov(x, t)$ at a single point in time and space. Since we are not interested in the variance along individual dimensions, we employ the generalized variance~\cite{Wilks1960, Wilks1967} defined as $|\det(\Cov(x, t))|$. Intuitively, this measures the multidimensional scatter of a Gaussian. To enable comparisons in different dimensions, we standardize this quantity by taking the $n$-th root in $n$-dimensional space. Lastly, we average this measure over time during the material transport:
\begin{equation}
	\tv(x_0, t_0, t_1) := \frac{1}{|t_1 - t_0|} \int_{t_0}^{t_1} |\det(\Cov(x, t))|^{\frac{1}{\dimension}} \dd t,
\end{equation}
where $x = \flowmap(x_0, t_0, t)$.
%This results in a scalar field that can be effectively visualized, cf.~\autoref{fig:DG:Reduced} (c).

\section{Results}
\label{sec:Results}

In this section, we visualize the uncertain transport in a synthetic and a real-world dataset. Additional results\revi{,} datasets\revi{, as well as exemplary source code} can be found in the supplementary material.

\subsection{Double Gyre}
\label{sec:Results:DoubleGyre}

\newlength{\dgheight}
\setlength{\dgheight}{79.0pt}

\begin{figure*}[t]
	\centering
	\subfloat[Density of advected particles]{%
		\includegraphics[height=\dgheight]{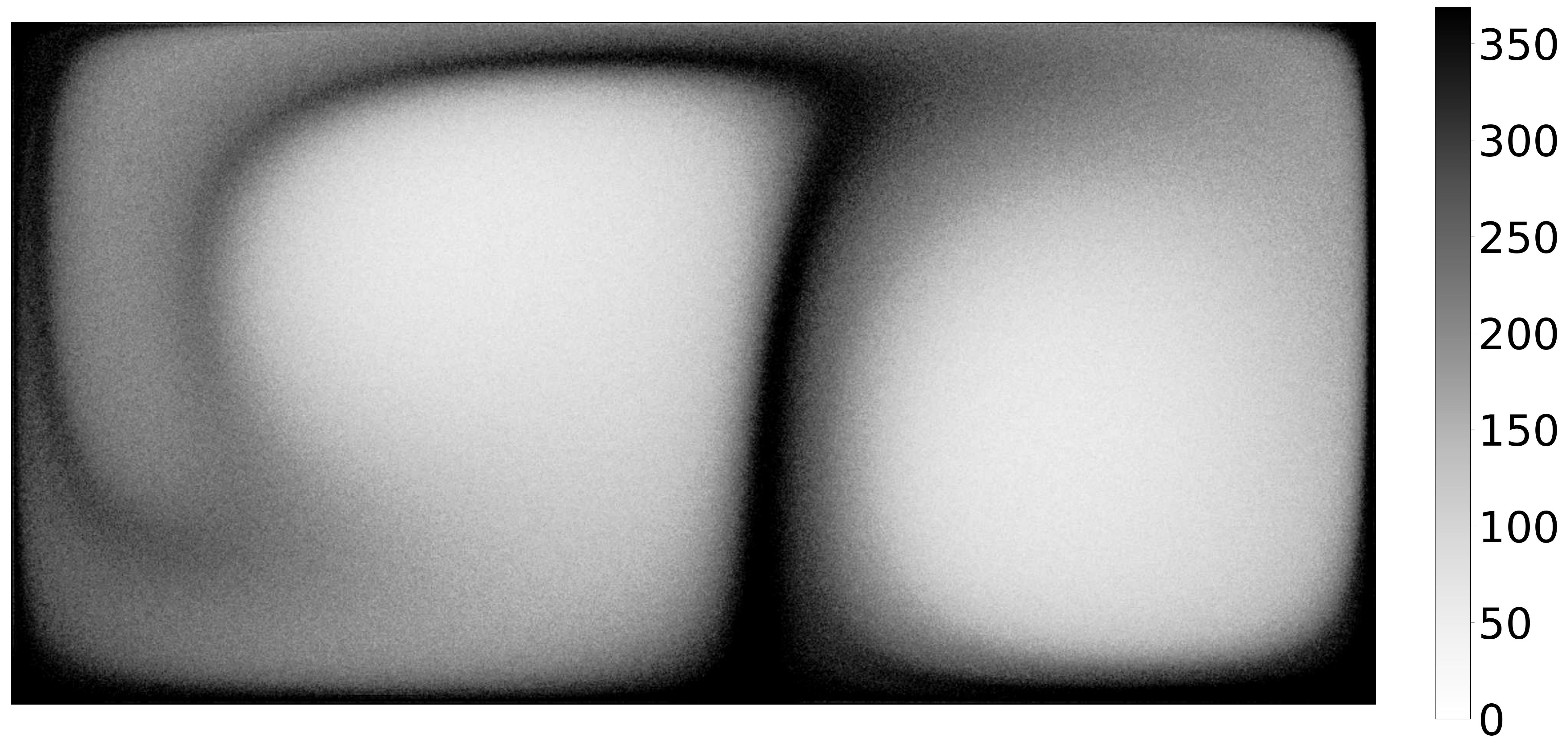}
	}
	\hfill
	\subfloat[DBS]{%
		\includegraphics[height=\dgheight]{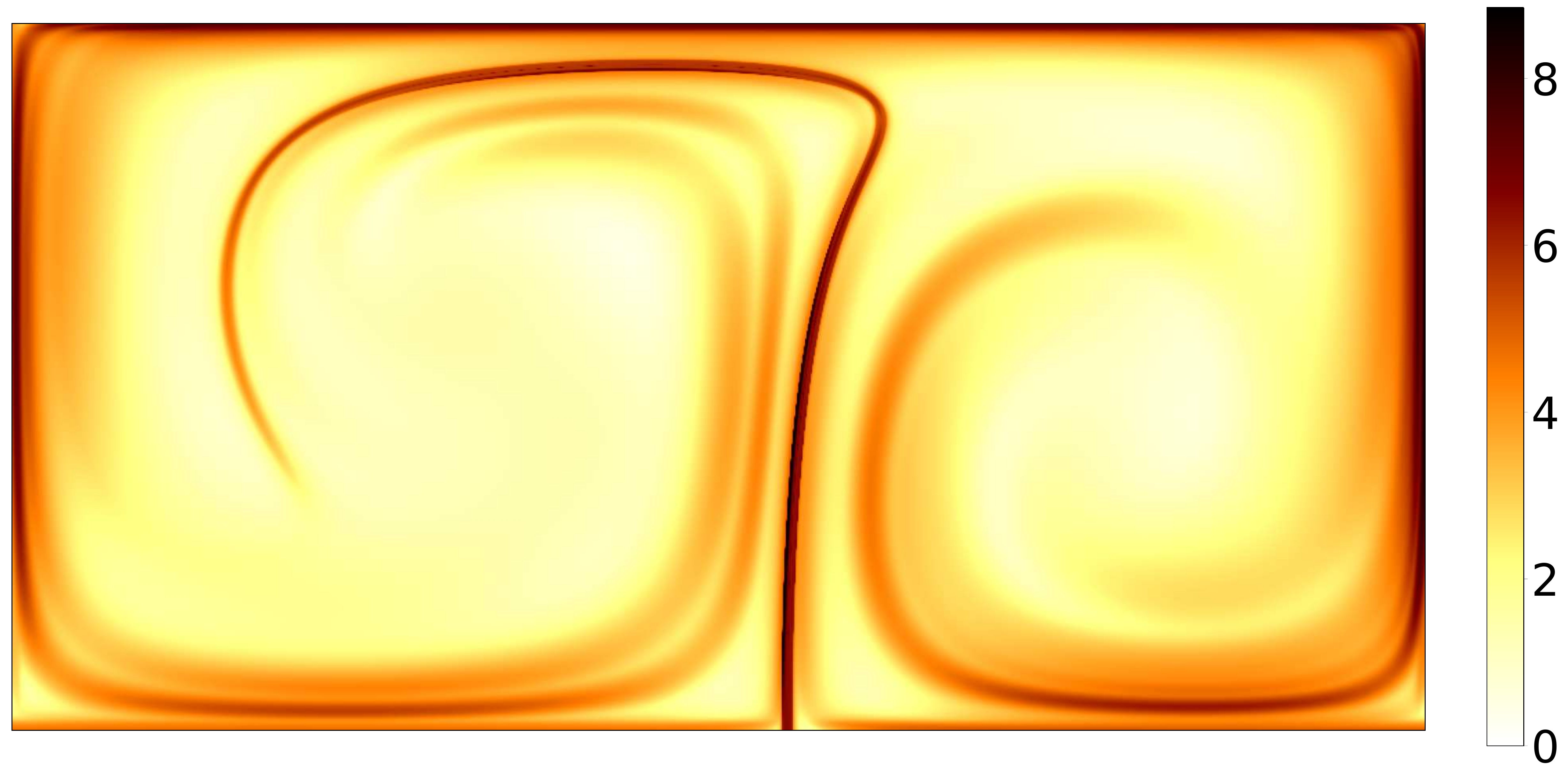}
	}
	\hfill
	\subfloat[Transport uncertainty]{%
		\includegraphics[height=\dgheight]{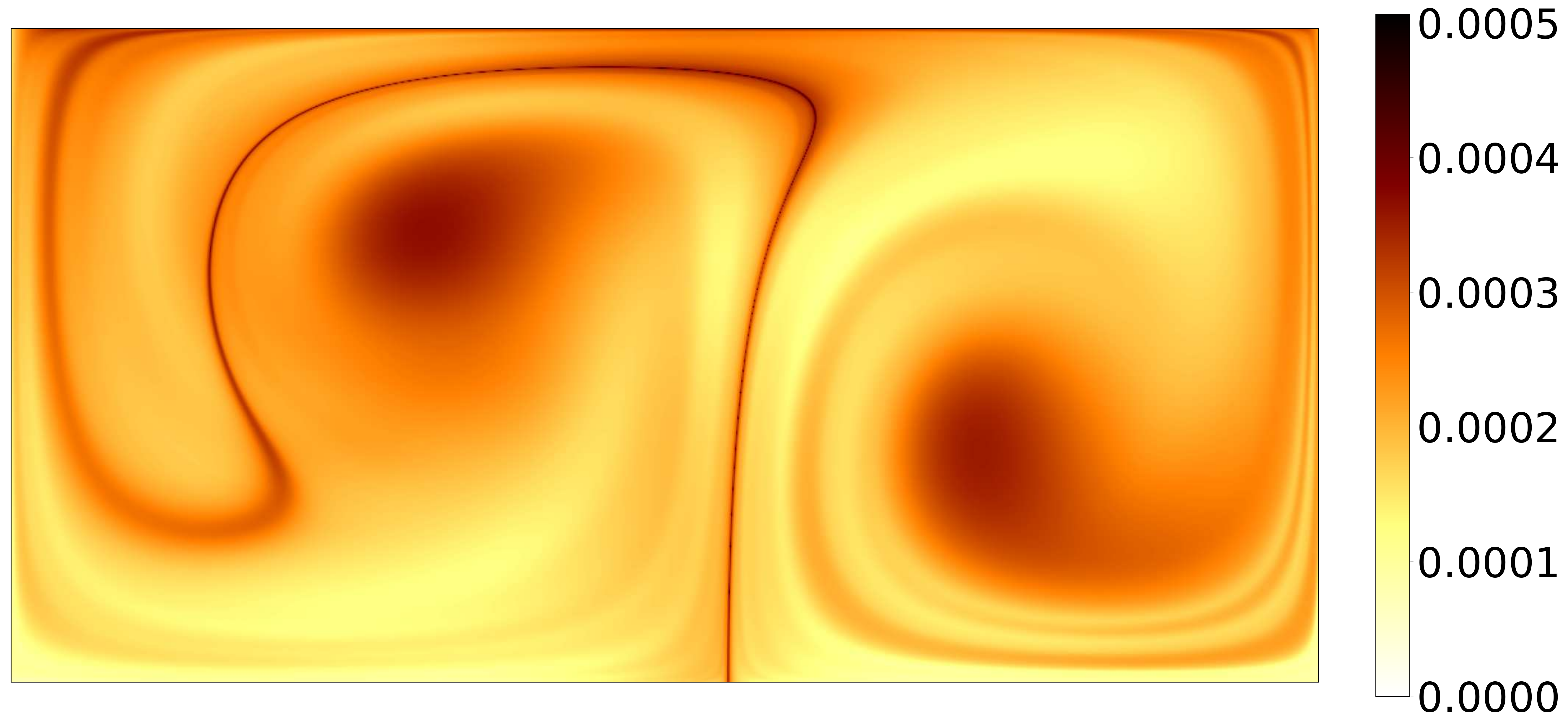}
	}
	\\
	\vspace*{-2mm}
	\subfloat[FTLE]{%
		\includegraphics[height=\dgheight]{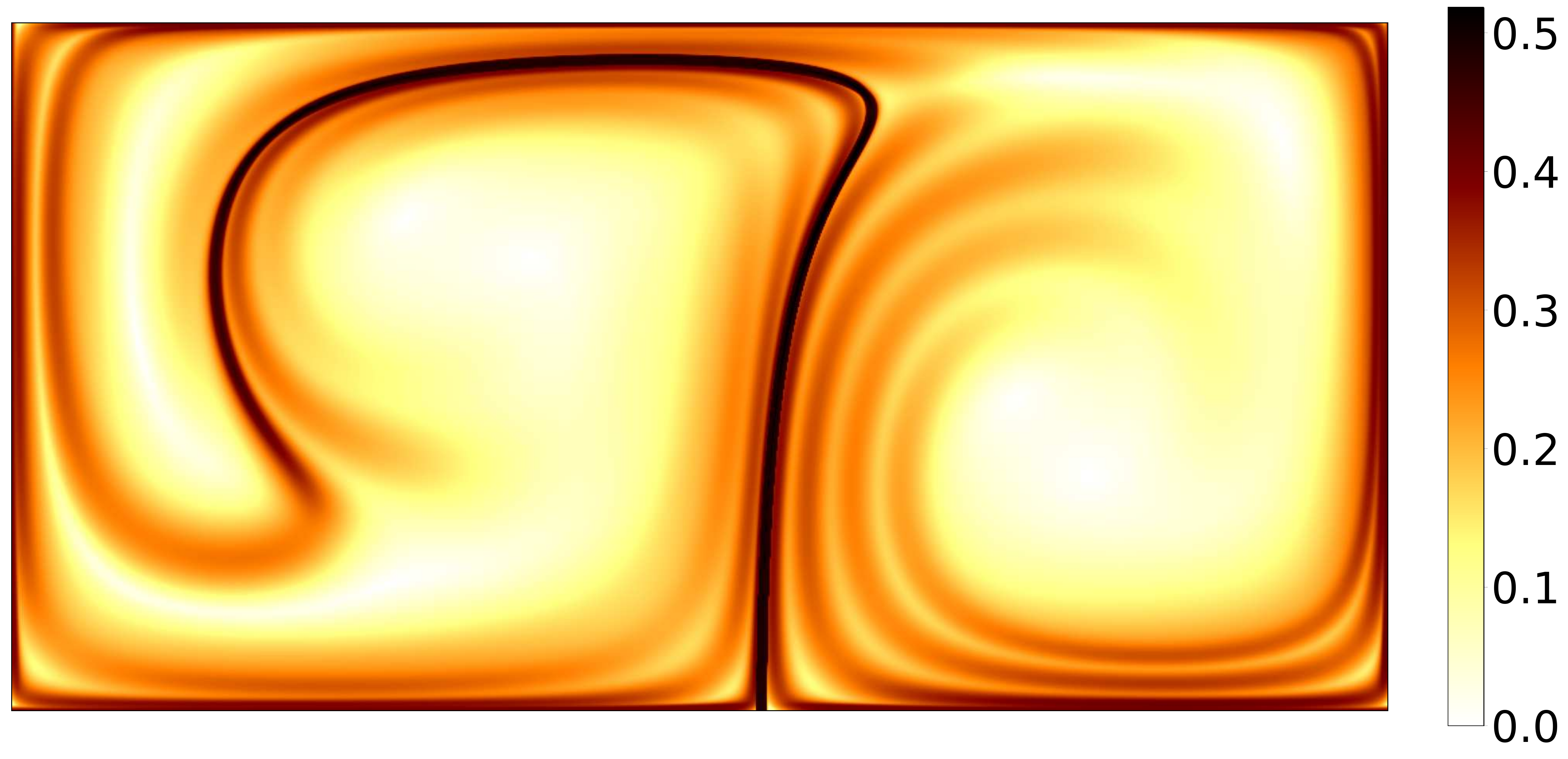}
	}
	\hfill
	\subfloat[D-FTLE (mean)]{%
		\includegraphics[height=\dgheight]{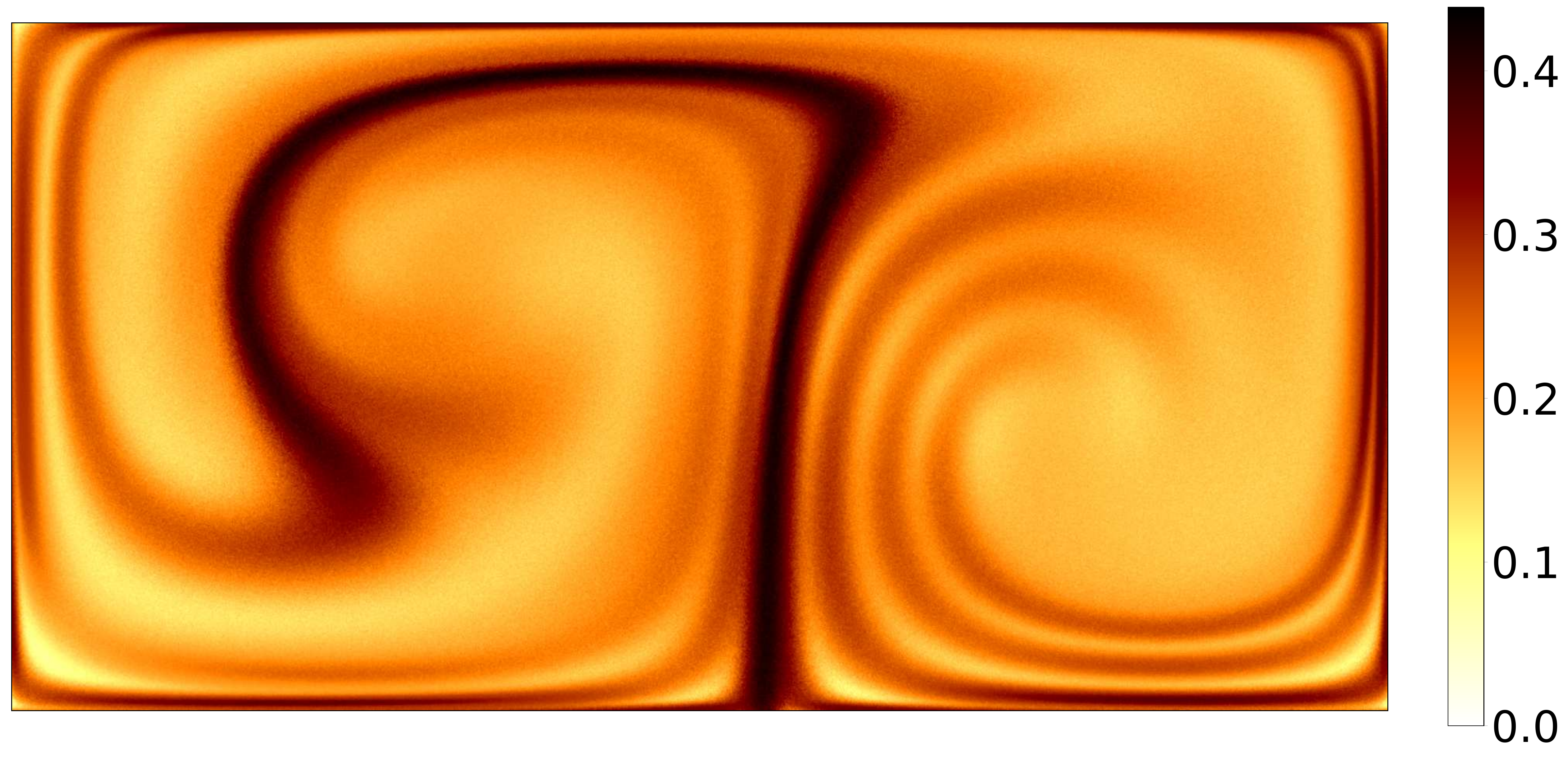}
	}
	\hfill
	\subfloat[D-FTLE (variance)]{%
		\includegraphics[height=\dgheight]{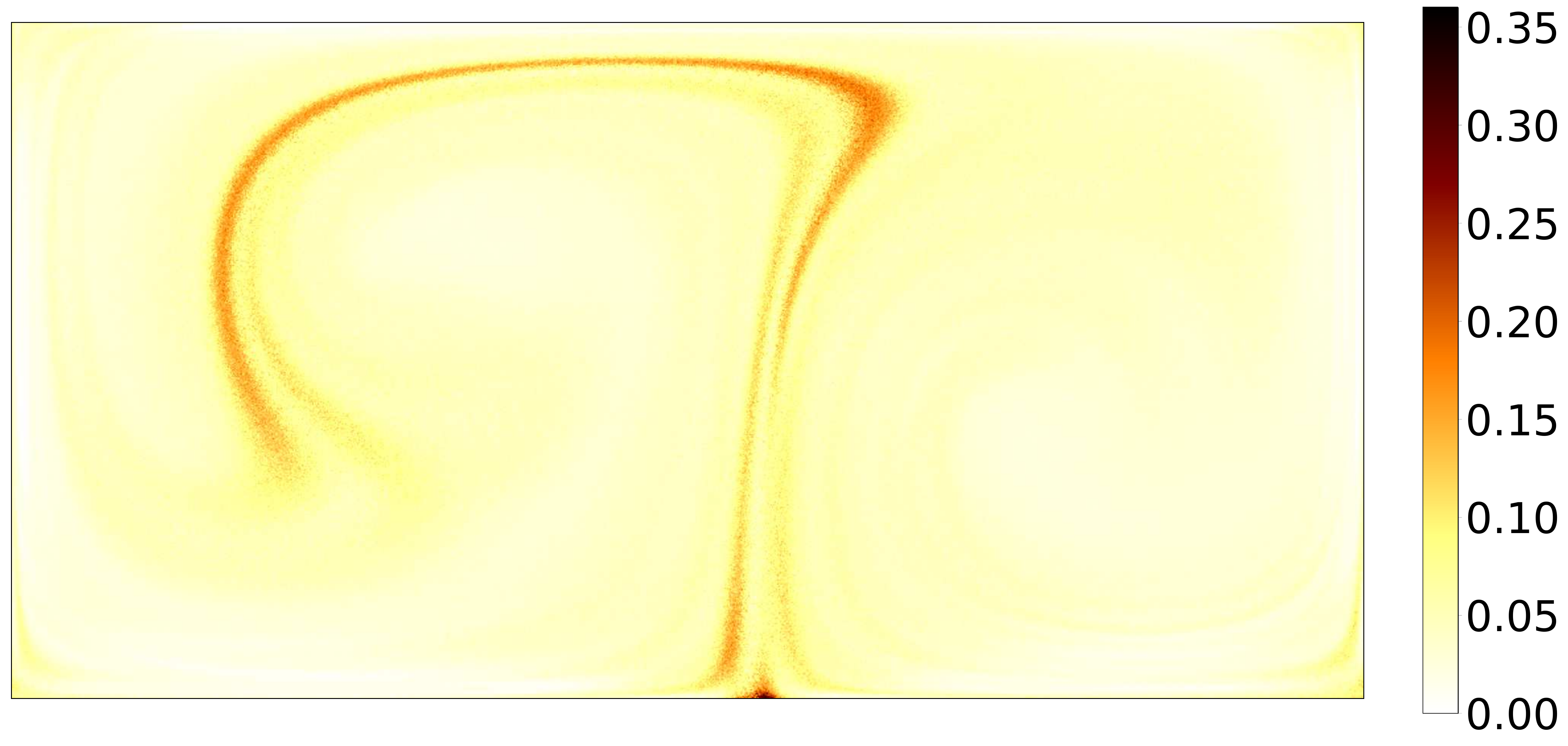}
	}
	\vspace*{-2mm}
	\caption{The Double Gyre dataset with uncertainty estimated during data reduction to a grid of size $[256 \times 128 \times 10]$.}
	\label{fig:DG:Reduced}
	\vspace*{-5mm}
\end{figure*}

This two-dimensional synthetic and time-dependent vector field describes two counter-rotating gyres. It is commonly used for the validation of FTLE and LCS\@. In the supplementary material, we describe the definition of the Double Gyre flow and perform additional experiments. In the following, we study the time interval $[0, 10]$ and integrate forward in time.

In \autoref{fig:DG:Reduced}, we employ a Gaussian error model estimated during data reduction to a space-time grid of size $[256 \times 128 \times 10]$.
In (a), we have stochastically advected a large amount of randomly distributed particles to visualize separating manifolds in the flow. The DBS shown in (b) clearly corresponds to these structures. The uncertainty of the transport is visualized in (c) and indicates a high uncertainty in the midst of both gyres. The DBS is low in these regions.
For reference, we illustrate the FTLE of the mean flow in (d), which does not consider the stochastic component of the flow. The FTLE indicates the presence of several smaller features around the two gyres that are not depicted in the density visualization in (a) or the DBS (b) and are located in regions of high uncertainty (c).

The mean D-FTLE from Guo et al.~\cite{Guo2016} shown in (e) indicates a larger amount of structures. The center of the left gyre even shows additional structures that are not present in the FTLE or the advected particles (a). The transport uncertainty indicates a high uncertainty in this area (c). 
However, the variance of the D-FTLE (f) is only high near the central barrier of the flow. At the same time, the presence of this barrier is far from uncertain. A high variance in the D-FTLE thus does not necessarily imply uncertainty of the transport barriers.

\subsection{Red Sea}

\begin{figure}
	\centering
	\subfloat[Temperature $t_0$]{%
		\includegraphics[width=0.495\linewidth]{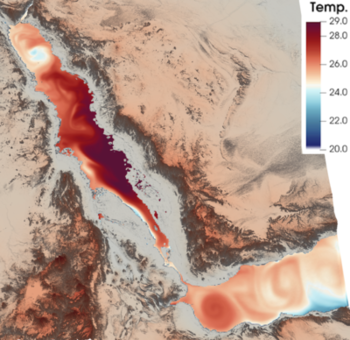}
	}
	\hfill
	\subfloat[Temperature $t_1$]{%
		\includegraphics[width=0.495\linewidth]{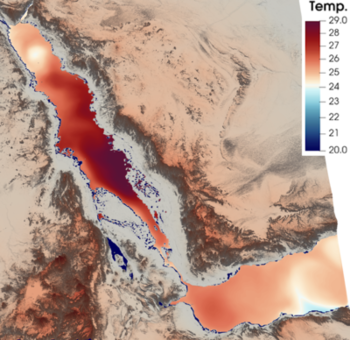}%
	}
	\caption{
	The mean temperature distribution in $t_0$ (a) and $t_1$ (b) illustrates the diffusion of temperature in the Red Sea and is closely aligned with the stochastic enhancers and barriers characterized by the DBS.}
	\label{fig:RedSea:Temperature}
	\vspace*{-4mm}
\end{figure}

This dataset from the SciVis contest 2020 is an ensemble simulation of the circulation dynamics in the Red Sea~\cite{Sanikommu2020}. Eddies in the ocean play a major role in the transport of energy and particles. Uncertainty is estimated from $50$ ensemble members created from perturbed initial conditions. Here, we estimate the mean and covariance from the individual members and analyze the resulting uncertain flow. 
Since the depth of the dataset is irregularly spaced, we have resampled it to a grid of size $500\times500\times150$ with $60$ time steps. To aid the understanding of the dataset, we have added topography and bathymetry~\cite{GEBCO} to our visualizations.

In \autoref{fig:teaser} (a) the backward DBS over a time interval of \num{182} hours is shown and indicates enhancers to stochastic transport. In \autoref{fig:RedSea:Temperature} (a) and (b) we visualize the diffusion of the temperature over time near the surface. Note that this diffusion corresponds to the enhancers indicated by the DBS.
The transport uncertainty shown in \autoref{fig:teaser} (b) indicates a high uncertainty in the gulf of Aden, the lower right part of the dataset. This suggests that we should investigate this region in more detail, for example by looking at individual ensemble members. Note that the DBS visualizes only the transport of the aggregated stochastic flow, but does not consider individual ensemble members.
%Lastly, a prior clustering of the ensemble members could be performed if the Gaussian assumption of the ensemble members is considered problematic, which was not the case for the Red Sea dataset.

% and assumes only small-scale deviations.
%Although this is a limitation, visualizing unsteady flows with large uncertainties is generally difficult if not infeasible due non-linear flow dynamics.

\subsection{Performance}

All of our evaluations were performed using GPU acceleration on an NVIDIA Quadro RTX 8000 with CUDA\@. To integrate deterministic flows, a fourth-order Runge-Kutta scheme is used. Stochastic flows are integrated using the Euler-Maruyama method with a constant number of \num{100} Monte Carlo runs.
%Note that we do not apply any of the optimization for the stochastic integration that Guo et al.~\cite{Guo2019} propose.

Performance measurements for different datasets are shown in \autoref{tbl:Performance}.
The DBS requires evaluating \autoref{eq:inv_transport_tensor} during each integration step. Computing the DBS thus takes two to four times longer than the FTLE.
In comparison, methods that depend on stochastic integration
increase the runtime by several orders of magnitude.

\begin{table}
	\centering
	\caption{Performance measurements of our datasets.}
	\vspace*{-2mm}
	\label{tbl:Performance}
	\resizebox{\linewidth}{!}{
		\begin{tabular}{lcrrr}
			\toprule
			Dataset & Resolution & FTLE & DBS & FTLE-D \\
			\midrule
			
			Double Gyre & $1024 \times 512$ & \ms{4.1} & \ms{7.4} & \ms{2173} \\
			
			Red Sea & $1500^2 \times 150$ & \ms{2739} & \ms{5193} & \ms{2129030} \\
			
			Heated Cylinder & $600 \times 1800$ & \ms{258.7} & \ms{680.0} & \ms{143243} \\
			
			Flow Around Corners & $2250 \times 750$ & \ms{110.8} & \ms{456.2} & \ms{220309} \\
			
			\bottomrule
		\end{tabular}
	}
	\vspace*{-4mm}
\end{table}

\subsection{Discussion}

Our results show that the DBS is significantly faster to compute than probabilistic extensions of the \revi{FTLE} since no stochastic integration is performed. In our experiments, the DBS closely aligns with the density of stochastically advected particles, whilst many features from the FTLE and \revi{its probabilistic extensions} are not visible. \revi{In fact, probabilistic extensions of the FTLE show features of possible realizations of an uncertain flow. In contrast, the DBS indicates features that exist in an inherently stochastic flow. Although both approaches have merit, it makes} the probabilistic D-FTLE hard to interpret.
At first glance, the variance of the D-FTLE might suggest uncertainties of the transport barriers and enhancers, however, this is not the case. Indeed, none of the approaches convey the actual amount of uncertainty encountered during integration. Our visualization of the transport uncertainties efficiently illustrates this uncertainty in the Lagrangian frame, thus providing additional insights.
%This is particularly important for the DBS, which is based on the assumption of small-scale deviations.

\section{Conclusion}% \& Future Work}

In this study, we introduce the theory of stochastic transport barriers and enhancers to the visualization community and discuss its application to Gaussian flow fields. %diffusion barrier strength 
%to visualize transport in uncertain flows.
The diffusion barrier strength, a quantity similar to the FTLE, visualizes the transport behavior in uncertain and unsteady flows.
%Similar to the theory of Lagrangian coherent structures and the FTLE, the diffusion barrier strength enables the analysis of uncertain flows.
% uncertain transport and mixing behavior.
%Due to its similarity to the FTLE, ridge extraction as well as most computational optimizations are directly applicable.
%
Compared to probabilistic extensions of the FTLE, the computation is significantly more efficient since no stochastic integration is performed. Moreover, the visualization is considerably simplified since the results are not probabilistic or averaged. By design, the DBS does not consider the absolute scale of deviations, which would require stochastic integration.
This assumption is problematic if regions of strong uncertainties exists in the flow. To this end, we propose a complementary visualization of the transport uncertainties that measures generalized variance in the Lagrangian frame of reference.
%
%In the future, we would like to relax the Gaussian assumption for ensemble datasets and consider clusters and outliers of the ensemble members.

%This limits the use of the DBS to flows with small uncertainties.
%To address this limitation in the future, we would like to cluster ensemble members to visually compare differences and outliers in the transport and mixing behavior.
%Lastly, the DBS can be applied to deterministic flows by assuming isotropic deviations, which enables its application together or even instead of the FTLE\@.

%% if specified like this the section will be committed in review mode
%\acknowledgments{
%The authors wish to thank A, B, and C. This work was supported in part by
%a grant from XYZ.}

%\bibliographystyle{abbrv}
\bibliographystyle{abbrv-doi}

\bibliography{../References}

\end{document}